\documentclass[aps,prb,amsmath,amssymb,twocolumn,superscriptaddress,showpacs,floatfix]{revtex4-1}
\usepackage{ifpdf}
 \newif\ifpdf
\ifx\pdfoutput\undefined
   \pdffalse
\else
   \pdfoutput=1
   \pdftrue
\fi
\ifpdf
   \usepackage{graphicx}
   \usepackage{epstopdf}
   \usepackage{color}
\else
   \usepackage{graphicx}
\fi
\usepackage{dcolumn}
\usepackage{bm}
\usepackage{srctex}
\usepackage{hyperref}

\begin{document}


\title{"Hall effect" for neutrons scattered by an A phase MnSi crystal}

\author{O.~G.~Udalov}
\affiliation{Institute for Physics of Microstructures, Russian Academy of Science, Nizhny Novgorod, 603950, Russia}
\affiliation{Department of Physics and Astronomy, California State University Northridge, Northridge, CA 91330, USA}

\author{A.~A.~Fraerman}
\affiliation{Institute for Physics of Microstructures, Russian Academy of Science, Nizhny Novgorod, 603950, Russia}
\affiliation{Department of physics and nanoelectronics, Lobachevsky State University of Nizhny Novgorod,
Nizhny Novgorod, Gagarin Avenue, 23, 603950, Russia}

\date{\today}


\begin{abstract}
We study a neutron diffraction by A phase of MnSi using a dynamical theory of diffraction and three wave approximation. We show that the neutron diffraction is asymmetrical with respect to an incident plane. The asymmetry depends on a sign of an external magnetic field. This phenomenon can be considered as the Hall effect for neutrons.
\end{abstract}

\maketitle

Currently the A phase of MnSi crystal attracts much scientific attention due to unusual properties of this phase. One of the interesting transport phenomena occurring in this crystal is the topological Hall effect (THE)~\cite{Pfleiderer2013,Pfleiderer2009Sc,Tokura2013,Chien2012,Tokura2012,Tokura2011}. This effect is the additional contribution to the Hall conductivity caused by an exchange interaction of conduction electrons with an inhomogeneous spin texture of MnSi Skyrmion lattice~\cite{Pfleiderer2009exp}. The exchange interaction leading to THE, can be described by the Pauli Hamiltonian $-J(\hat{\vec{\sigma}}\cdot \vec{m})$~\cite{Vonsovsky1974}, where $J$ is the coupling constant, $\hat{\vec{\sigma}}$ is the Pauli matrices, and $\vec{m}$ is the local magnetization. It is know that the interaction of a neutron with a magnetic field is defined by the same Hamiltonian~\cite{Gurevich}, $-(\hat{\vec{\mu}}_{n}\cdot \vec{B})$, where $\hat{\vec{\mu}}_{n}$ is the neutron magnetic moment operator, and $\vec{B}$ is the induction of a magnetic field. One can expect the appearance of topological Hall phenomenon for neutrons scattered by the Skyrmion lattice due to the above mentioned interactions' similarity. In contrast to the electron's exchange coupling the magneto-dipole interaction of neutrons is rather weak. Therefore, the quasiadiabatic approximation~\cite{Aharonov} used for derivation of THE, is not valid for neutrons. However, it was recently demonstrated that the  Pauli term can induce the Hall effect even in the limit opposite to the quasiadibatic approximation~\cite{Ud2013}. It was shown that the neutron beam, diffracted by a ferromagnetic crystal, has the intensity which depends on the sign of the crystal magnetization and contains the term $(\vec{M}\cdot [\vec{k}\times \vec{k}'])$, where $\vec{M}$ is the crystal magnetization, and vectors $\vec{k}$ and $\vec{k}'$ are the wavevectors of incident and diffracted neutrons. The effect was called "skew" scattering of unpolarized neutrons. It can be observed in specific scattering geometry when the Bragg condition is satisfied for two reciprocal vectors simultaneously. In the present paper we demonstrate that the Hall effect appears for neutrons diffracted by the Skyrmion lattice of a MnSi crystal.

\begin{figure}[h]
\includegraphics[width=0.6\columnwidth, keepaspectratio=true]{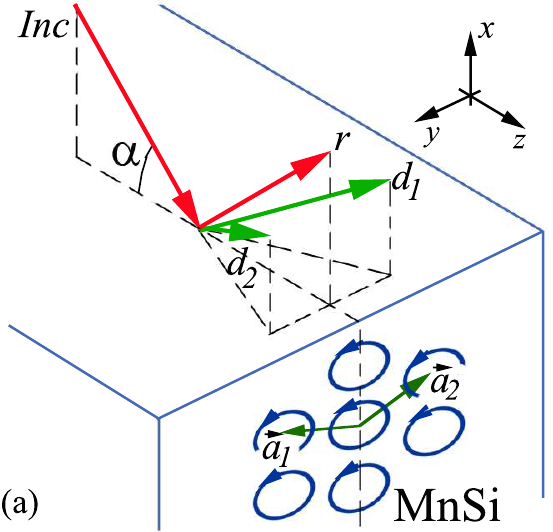} \\ \includegraphics[width=0.5\columnwidth, keepaspectratio=true]{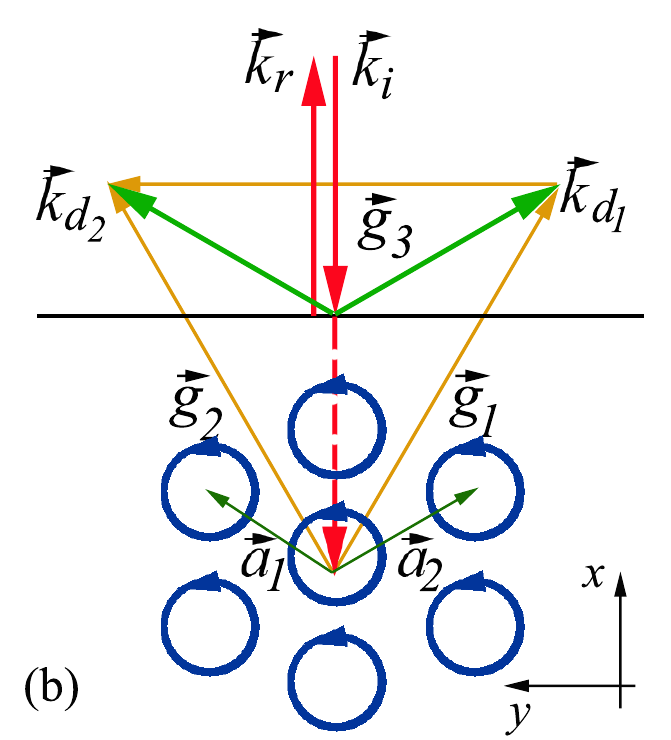}
\caption{\label{Fig1} The neutron diffraction geometry. The subscripts $i$, $r$, $d_1$, and $d_2$ denote incident, reflected, and two diffracted beams. Unpolarized neutrons fall on a MnSi surface with a glancing angle $\alpha$. The angle $\beta$ determines the orientation of the incident plane with respect to (x,z) plane. Rings with arrows symbolize a Skyrmion lattice with lattice vectors $\vec{a}_1$, and $\vec{a}_2$. Corresponding reciprocal vectors are $\vec{g}_1$, $\vec{g}_2$, and $\vec{g}_3$. These vectors lie in the plane (y,x). Figure (a) depicts the general view of the system. Figure (b) shows the z-direction view. Figure (b) corresponds to the most symmetric case when $\beta=0$.}
\end{figure}

The effect similar to the neutron "skew" scattering was predicted in Ref.\cite{MaleevToperverg1978}, where it was shown that an inelastic scattering cross section of polarized neutrons contains the following contribution $(\vec{P}\cdot[\vec{k}\times \vec{k}'])$, where $\vec{P}$  is the spin polarization of an incident neutron beam. This effect was observed experimentally in Ref.~\cite{OkorokovGusakov1978}. We consider the case of elastic scattering of unpolarized neutrons in contrast to Ref~.\cite{MaleevToperverg1978}.

Many neutrons diffraction experiments on MnSi were performed recently~\cite{Monch2012,Rosch2011,Grig2006,Ishi1983}. The experimental geometry, used in Ref.~\cite{Pfleiderer2009Sc}, is almost the same as we discuss in the present paper (see Fig.~\ref{Fig1}). In this geometry two reciprocal vectors satisfy the Bragg condition. Thus the necessary circumstance for the "skew" neutron scattering was realized in the experiments of Ref~\cite{Pfleiderer2009Sc}. However, rather high neutron beam divergence did not allow observing the Hall effect for neutrons. Below we discuss the results of Ref.~\cite{Pfleiderer2009Sc} from the point of view of out theoretical prediction.

In the present work we consider the geometry shown in Fig. 1 where a MnSi crystal is in the A phase.  An external magnetic field is applied along the z axis with vortices (Skyrmions) lines being co-directed with this field. Neutrons with the wavevector $\vec{k}_i$ fall on the crystal at an incident angle $\alpha$. The neutron incident plane makes an angle $\beta$ with the (x,z) plane. Two diffracted beams come out from the crystal with the wavevectors $\vec{k}_{d_{1,2}}$. One of the Skyrmion lattice vectors is parallel to the x axis. Due to a sixfold axis of symmetry, there are two other lattice vectors aligned symmetrically with respect to the (x,z) plane in directions ($1/2,\pm \sqrt{3}/2,0$). These vectors are denoted as $\vec{a}_{1,2}$ in Fig.~\ref{Fig1} . The MnSi magnetic structure has a period $|\vec{a}_{1,2}|=20.78$ nm~\cite{Pfleiderer2009exp} with reciprocal lattice vectors $\vec{g}_{1,2,3}$ being aligned along ($0,1,0$) and ($\sqrt{3}/2,\pm 1/2,0$) directions (see Fig.~\ref{Fig1}(b)). These vectors have equal magnitudes $g=|\vec{g}_{1,2,3}|=0.349$ nm$^{-1}$. To satisfy the Bragg condition  $|\vec{k}_i+\vec{g}_{1,2}|\approx|\vec{k}_i|$, the perpendicular (x,y) component of the incident neutron wave vector $\xi_0=\sqrt{(k_i^x)^2+(k_i^y)^2}$ has to be approximately equal to $g/\sqrt{3}$. At small enough $\beta$ the Bragg condition is satisfied for two reciprocal vectors simultaneously. For neutrons with wavelength of 1 nm the glancing angle of incidence is  $\alpha\approx\alpha_0=1.842^\circ$.

The neutron reflection from the MnSi crystal is rather weak. Therefore a sample surface  orientation is almost unimportant in the neutron diffraction experiment on MnSi. The most interesting phenomena appearing in the system are due to the waves interference in the bulk. For simplicity we consider the case where MnSi surface is parallel to the (y,z) plane.

We find the intensity of unpolarized neutrons diffracted by the MnSi crystal using the dynamical theory of neutron diffraction. The theory was initially formulated in Ref.~\cite{Stassis1974}. Outside the crystal a neutron wave function consists of four plane waves describing incident, reflected, and two diffracted beams. These beams have the following wavevectors: $\vec{k}_i=$($-\xi_0,0,k_z$), $\vec{k}_r=$($\xi_0,0,k_z$), and $\vec{k}_{d_{1,2}}=$($\sqrt{\xi_0^2-g^2/4},\pm g/2, k_z$) correspondingly.
Inside a crystal there are twelve neutron waves with different quasimomentums and spins. These waves are described by the following system of equations

\begin{equation}\label{Eq_GenSys}
\left\{\begin{array}{l} {(1-\frac{k^{2}}{\xi_0^{2}} -\hat{V}_{0})\Psi _{0}
-\hat{V}_{-g_1}\Psi_{r} -\hat{V}_{-g_2}\Psi_{l} =0,} \\ {-\hat{V}_{g_1}\Psi _{0}
+(1-\frac{(\vec{k}+\vec{g}_{1})^{2}}{\xi_0^{2}} -\hat{V}_{0})\Psi_{r}
-\hat{V}_{-g_3}\Psi_{l} =0,} \\ {-\hat{V}_{g_2}\Psi_{0}-\hat{V}_{g_3}\Psi_{r}
+(1-\frac{(\vec{k}+\vec{g}_{2})^{2}}{\xi_0^{2}}-\hat{V}_0)\Psi_{l} =0.}
\end{array}\right.
\end{equation}
Here $\vec{k}$ is the neutron quasimomentum. A neutron state with a certain quasimomentum $\vec{k}$ is the combination of three plane waves with the wavevectors $\vec{k}$, $\vec{k}+\vec{g_1}$, and $\vec{k}+\vec{g_2}$. The amplitudes of these plane waves are denoted by $\Psi_0$, $\Psi_r$, and $\Psi_l$. They are two component spinors. Spin dependent operators $\hat{V}_0$, and $\hat{V}_{g_{1,2,3}}$ are the spatial Fourier harmonics of the potential acting on neutrons inside the crystal normalized to the quantity $\hbar^2\xi^2_0/2m_n$. $\hat{V}_0$ stands for a zero spatial frequency. We consider the neutron wavelength larger than the period of MnSi lattice, therefore the nuclear potential inside the crystal can be considered as uniform with only a zero frequency harmonic. Using data on the MnSi coherent scattering length, we estimate the nuclear potential $V_{nuc}\approx 1.478\cdot 10^{-27}$ J.

We now discuss the contribution of a neutron magneto-dipole interaction in potentials $\hat{V}_{0,g_{1,2,3}}$. Consider a magnetic unit cell of the Skyrmion lattice formed by the vectors $\vec{a}_1$ and $\vec{a}_2$. Due to a sixfold symmetry axis of the lattice the spatial distribution of magnetic induction $\vec{B}(\vec{r})$ (taking into account the external field) obeys the following relations: $B_z(-x,y,z)=B_z(x,y,z)=B_z(x,-y,z)$, $B_x(-x,y,z)=B_x(x,y,z)=-B_x(x,-y,z)$ and $-B_y(-x,y,z)=B_y(x,y,z)=B_y(x,-y,z)$. Using these equations and the sixfold rotational symmetry of the Skyrmion lattice one can show that the spatial Fourier harmonics of the magneto-dipole interaction potential have the form:
$\hat{V}^{md}_{0}=\mu_n(\vec{B}_0\cdot\hat{\vec{\sigma}})=\mu_n B_{z0}\hat{\sigma}_z$, $\hat{V}^{md}_{g_1}=\mu_n(\vec{B}_1\cdot\sigma)=\mu_n B_{z1}\hat{\sigma}_z+\mu_n B_{\perp}(1/2\hat{\sigma}_x+\sqrt{3}/2\hat{\sigma}_y)$,
$\hat{V}^{md}_{g_2}=\mu_n(\vec{B}_2\cdot\hat{\vec{\sigma}})=\mu_n B_{z1}\hat{\sigma}_z+\mu_n B_{\perp}(-1/2\hat{\sigma}_x+\sqrt{3}/2\hat{\sigma}_y)$,
and $\hat{V}^{md}_{g_3}=\mu_n(\vec{B}_3\cdot\hat{\vec{\sigma}})=\mu_n B_{z1}\hat{\sigma}_z-\mu_n B_{\perp}\hat{\sigma}_x$. Note that $\hat{V}^{md}_{0,g_1,g_2}$ should be normalized to get $\hat V_g$. The quantity $B_{\perp}$ has imaginary part only, since the distribution of magnetic induction in the (x,y) plane is antisymmetric. Therefore, the spatial distribution of magnetic induction in the lattice of Skyrmions is characterized by three parameters $B_{z0}$, $B_{z1}$, and $B_{\perp}$. These parameters are unknown, but can be found from an experiment. The value of Mn magnetic moment is $0.4\mu_B$, where $\mu_B$ is the Bohr magneton. Thus, the interaction energy of neutrons and magnetic field created by MnSi can be estimated as $E_m\approx 1.889\cdot 10^{-27}$ J.

The system of Eqs.~(\ref{Eq_GenSys}) generates 12 different neutron states inside the crystal with the same energy. Six of them correspond to neutrons going outward the MnSi surface and six describe neutrons propagating to the surface. We consider a semi-infinite crystal. Therefore, only the outgoing neutrons should be taken into account. Using boundary conditions~\cite{Stassis1974,Ud2013} the intensity of reflected and diffracted neutron waves can be evaluated.

We calculate the intensity of the neutrons diffracted to the left (wavevector $\vec{k}_{d_1}$) and to the right (wavevector $\vec{k}_{d_2}$) with respect to the incidence plane assuming the incident beam is unpolarized. The detailed description of procedure of calculations is given in the Appendix~\ref{Sec:App}. Fig.~\ref{Fig_Dif2D} shows the dependence of the diffracted beam intensity on the angles $\alpha$ and $\beta$. The dependencies are calculated for the following set of parameters: $\mu B_0=E_m$, $\mu B_z=E_m/2$, $\mu |B_{\perp}|=E_m/1.5$. Figures (a) and (b) correspond to the beams diffracted to the left ($\vec{k}_{d_1}$) and to the right ($\vec{k}_{d_2}$), correspondingly. Bright (orange online) color corresponds to high intensity. Dark (blue online) color shows low intensity regions. Diagonal bright (orange online) stripe shows the diffraction peak. The angle $\alpha$ at which the diffraction intensity reaches it's maximum depends on the angle $\beta$. It can be understood as follows. Diffraction peak position can be approximately determined by the condition $|\vec{k}_i+\vec{g}|=|\vec{k}_i|$. Vector $\vec{k}_{\perp}=(k_i^x, k_i^y,0)$ denotes the perpendicular component of the incident wavevector. It's magnitude can be found as $\xi_0=|\vec{k}_{\perp}|=|k_{i}|\sin(\alpha)$. Denote $\tan(\phi)=k_i^y/k_i^x$. For small $\beta$ the angle $\phi\approx\beta$. Bragg condition has the form $g^2+2(\vec k_\perp\cdot\vec g)=0$. One can find the angles $\alpha_{g_{1,2}}$ at which the diffraction peaks appear by the expression $\alpha_{g_{1,2}}\approx g^2/(2|\vec k_i|(g_x\pm g_y\beta))$. Thus the angle $\alpha$ at which diffraction peak appears, depends on the incident plane orientation $\beta$. Diffraction peaks $\vec{k}_{d_1}$ and $\vec{k}_{d_2}$ "move" in opposite directions with changing $\beta$.

\begin{figure}[h]
\includegraphics[width=1\columnwidth, keepaspectratio=true]{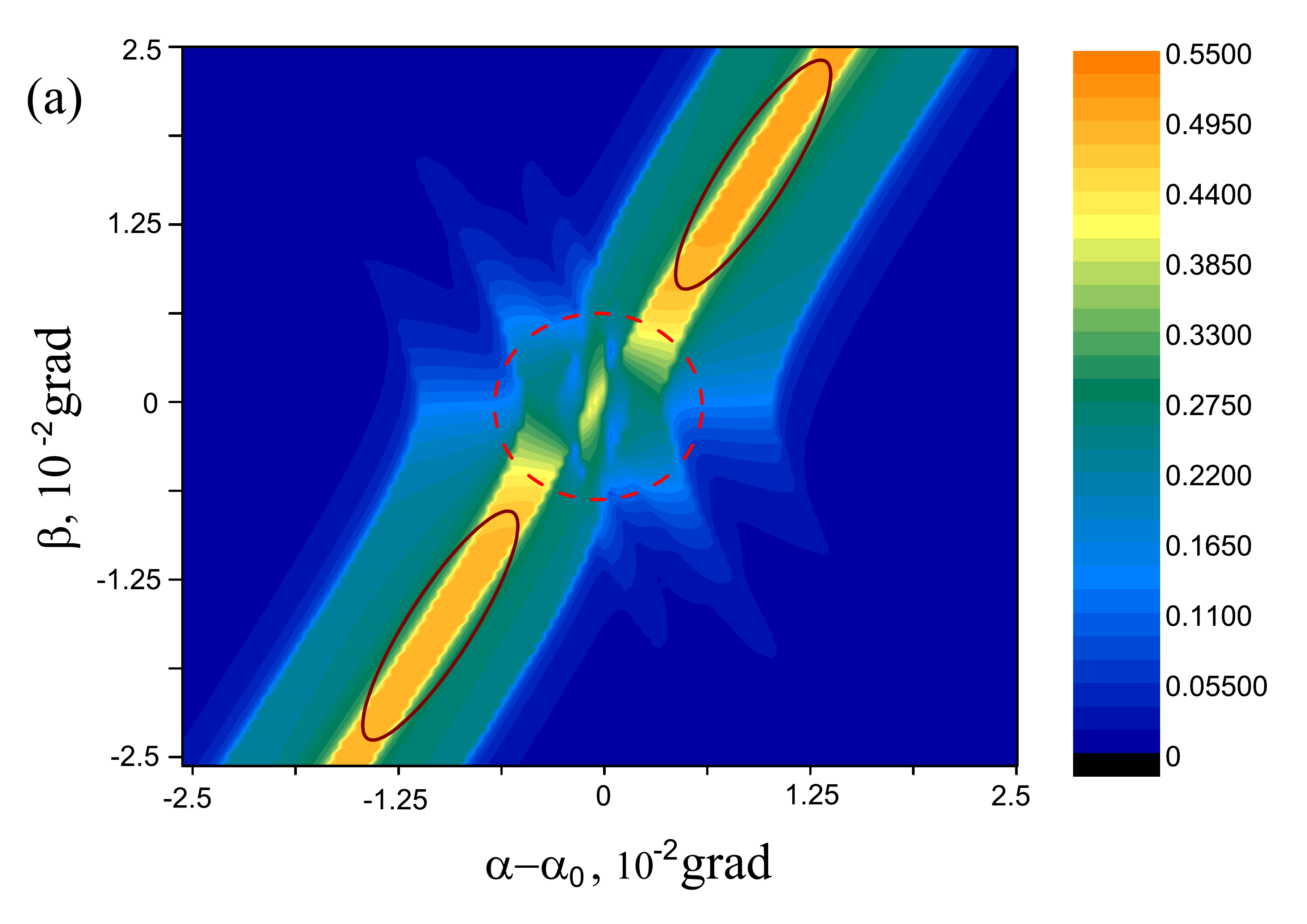} \\ \includegraphics[width=1\columnwidth, keepaspectratio=true]{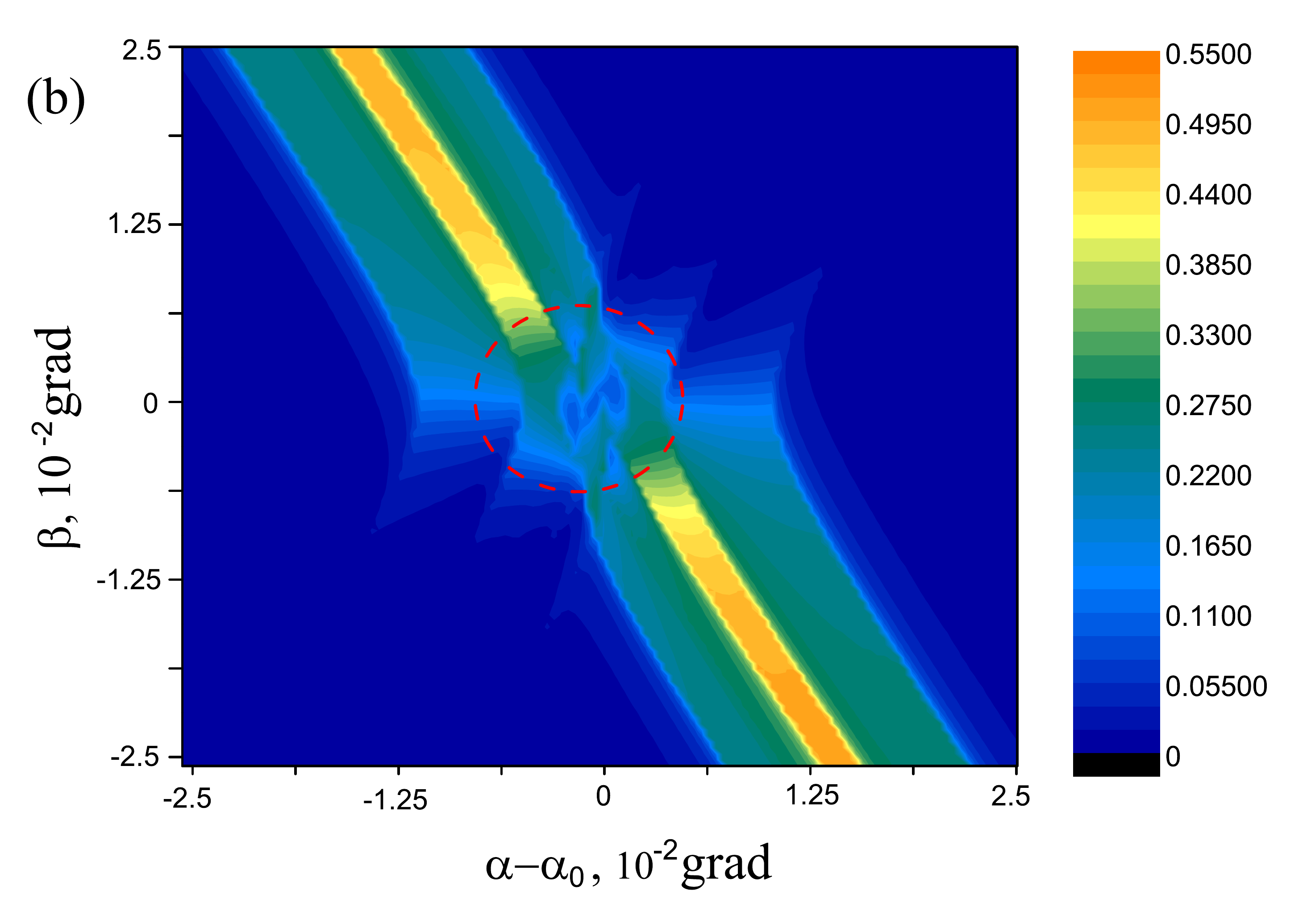}
\caption{\label{Fig_Dif2D} Diffracted intensity as a function of the angles $\alpha$ and $\beta$. $\alpha_0=1.842$ grad. (a) and (b) correspond to diffraction directions $\vec{k}_{d_1}$ and  $\vec{k}_{d_2}$. Ovals show the regions where only one reciprocal vector satisfies the Bragg condition. In these regions two wave approximation is valid. Diffracted intensity in these regions obey the relation $I_{d_1}(\alpha,\beta)=I_{d_2}(\alpha,-\beta)$. Circles show the regions in which two reciprocal vectors satisfy the Bragg condition and three wave approximation has to be used. In these regions skew scattering appears leading to difference between diffraction peaks $\vec{k}_{d_1}$ and  $\vec{k}_{d_2}$ and violation of the relation $I_{d_1}(\alpha,\beta)=I_{d_2}(\alpha,-\beta)$.}
\end{figure}

Ovals at Fig.~\ref{Fig_Dif2D}(a) show the regions in which only one reciprocal vector satisfies the Bragg condition. Neutron scattering in this regions can be approximately treated in the two wave approximation. Inside the crystal neutrons are described by the following equations~\cite{Stassis1974}

\begin{equation}\label{Eq_TwoWaveSyst}
\left\{\begin{array}{l} {(1-\frac{k^{2}}{\xi_0^{2}} -\mu_n(\hat{\sigma}\cdot\vec{B}_0))\Psi _{0}
-\mu_n(\hat{\sigma}\cdot\vec{B}_{-g})\Psi_{g}=0,} \\ {-\mu_n(\hat{\sigma}\cdot\vec{B}_{g})\Psi _{0}
+(1-\frac{(\vec{k}+\vec{g})^{2}}{\xi_0^{2}} -\mu_n(\hat{\sigma}\cdot\vec{B}_0)))\Psi_{g} =0.}
\end{array}\right.
\end{equation}
Here $\vec{g}$ is the lattice reciprocal vector (can be either $\vec{g}_1$ or $\vec g_2$). A wave function of a neutron with the quasimomentum $\vec{k}$ inside a crystal consists of two plane waves with amplitudes $\Psi_{0}$ and $\Psi_{g}$. These plane waves have wavevectors $\vec{k}$ and $\vec{k}+\vec{g}$. $\vec{B}_0$ and $\vec{B}_g$ are the Fourier coefficients of a magnetic induction spatial distribution inside the crystal. They correspond to harmonics with zero and $\vec{g}$ wavevectors.

\begin{figure}[h]
\includegraphics[width=1\columnwidth, keepaspectratio=true]{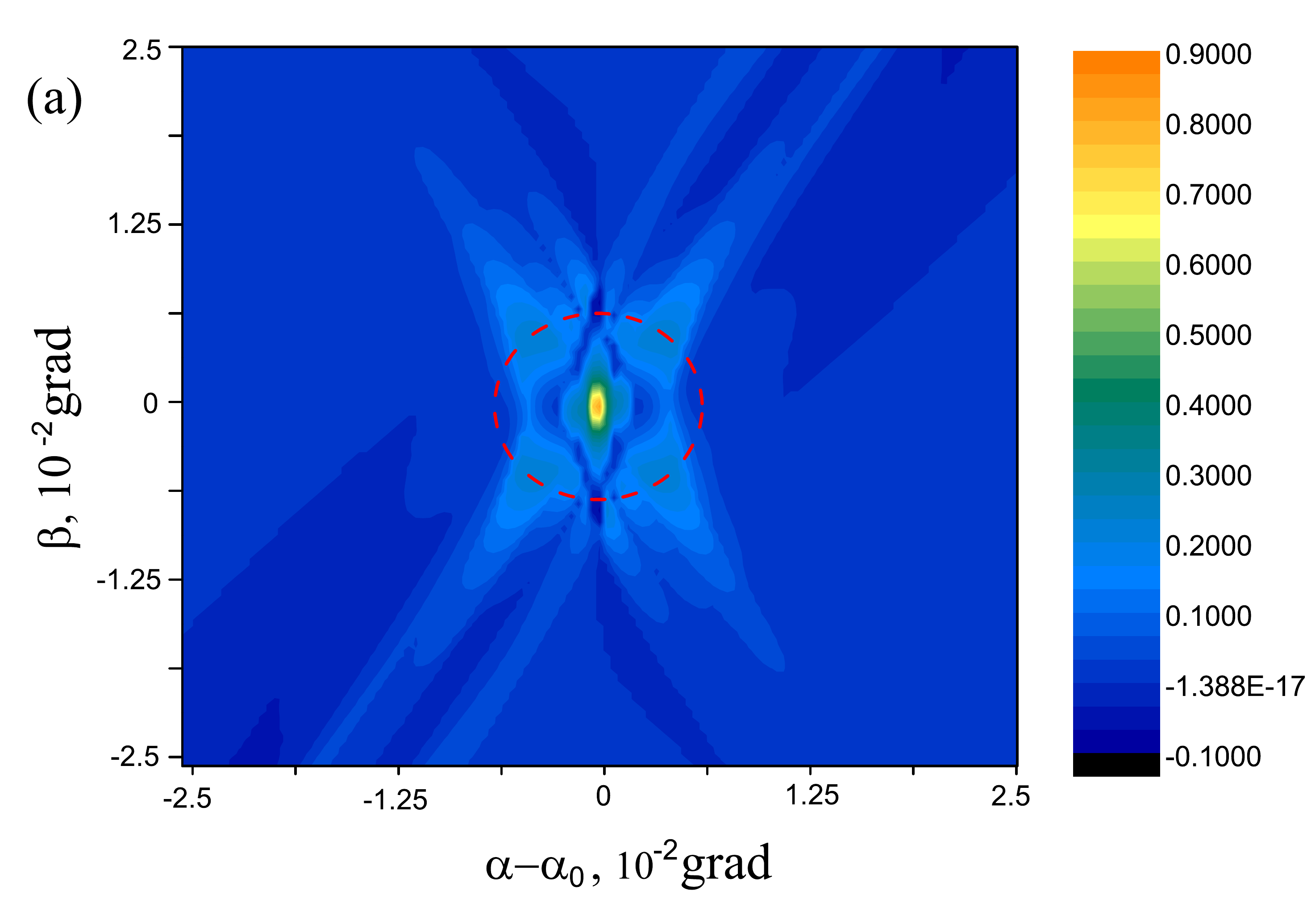} \\\includegraphics[width=1\columnwidth, keepaspectratio=true]{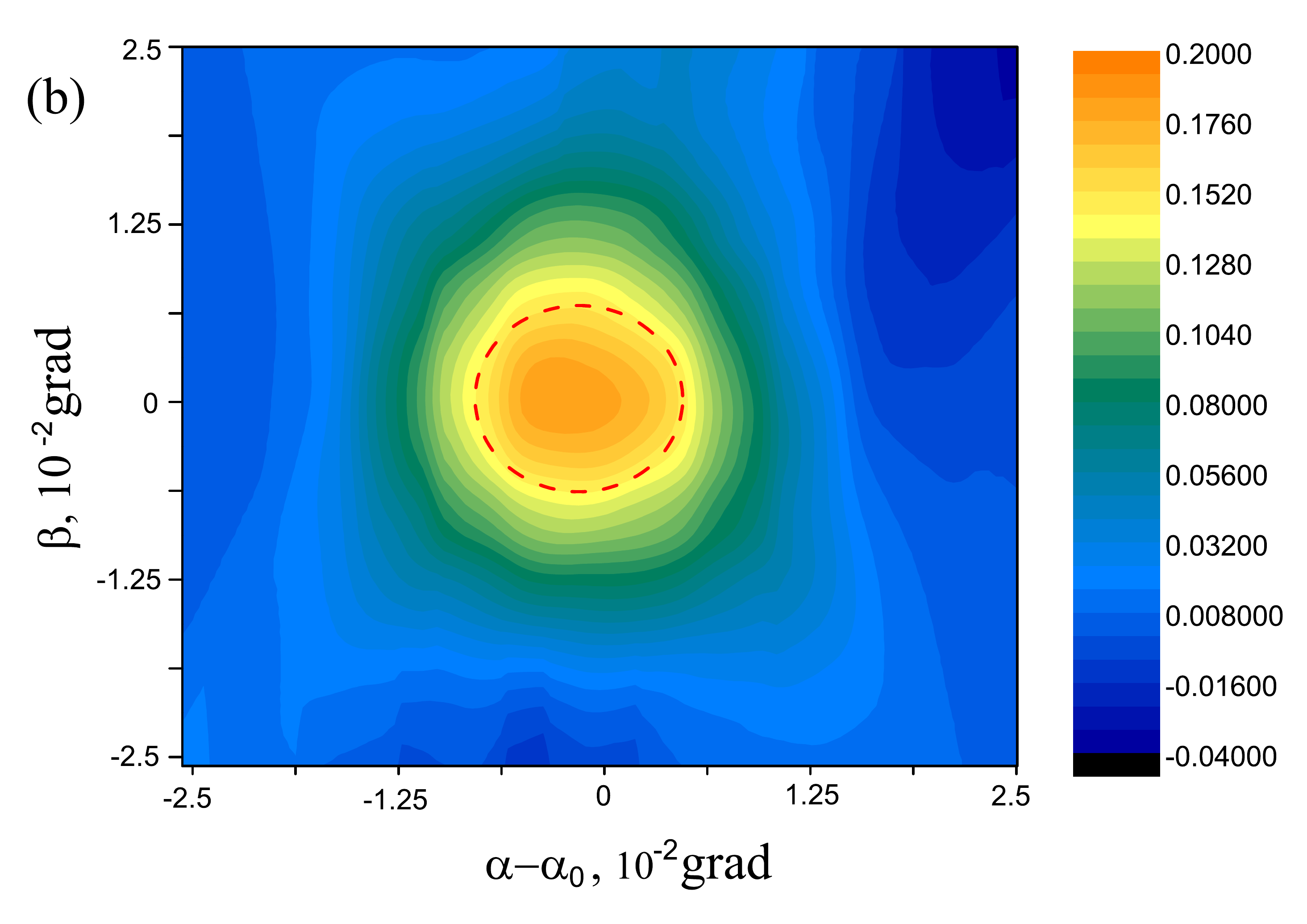}
\caption{\label{Fig_DeltaDif} (a) The quantity $2\Delta I(\alpha,\beta)/(I_{d_1}(\alpha_0,0)+I_{d_2}(\alpha_0,0))$ as a function of the angles $\alpha$ and $\beta$. (b) The dependence of the averaged quantity $2\Delta I(\alpha,\beta)/(I_{d_1}(\alpha_0,0)+I_{d_2}(\alpha_0,0))$ on the angles $\alpha$ and $\beta$.}
\end{figure}

These equations are invariant with respect to the following transformation $k_y\to-k_y$, $g_y\to -g_y$ ($\vec g_1\to\vec g_2$). Due to this symmetry the intensity of neutrons diffracted to the left and to the right from the (x,z) plane, obeys the following relation $I_{d_1}(\alpha,\beta)=I_{d_2}(\alpha,-\beta)$. Fig.~\ref{Fig_DeltaDif}(a) demonstrates the quantity $\Delta I(\alpha,\beta)=I_{d_1}(\alpha,\beta)-I_{d_2}(\alpha,-\beta)$ as a function of the angles $\alpha$ and $\beta$. The quantity $\Delta I(\alpha,\beta)$ is zero for all $\alpha$ and $\beta$ besides the central region ($\alpha\approx\alpha_0$, $\beta\approx0$) of the plot.

Central region of the plot (showed by a circle) corresponds to the case when two reciprocal vectors satisfy the Bragg condition. In this case three wave approximation (Eq.~\ref{Eq_GenSys}) should be used. The above mentioned symmetry of the diffraction peaks intensity is broken in this region $I_{d_1}(\alpha,\beta)\ne I_{d_2}(\alpha,-\beta)$. This is a signature of the neutron skew scattering.

\begin{figure}[t]
\includegraphics[width=0.95\columnwidth, keepaspectratio=true]{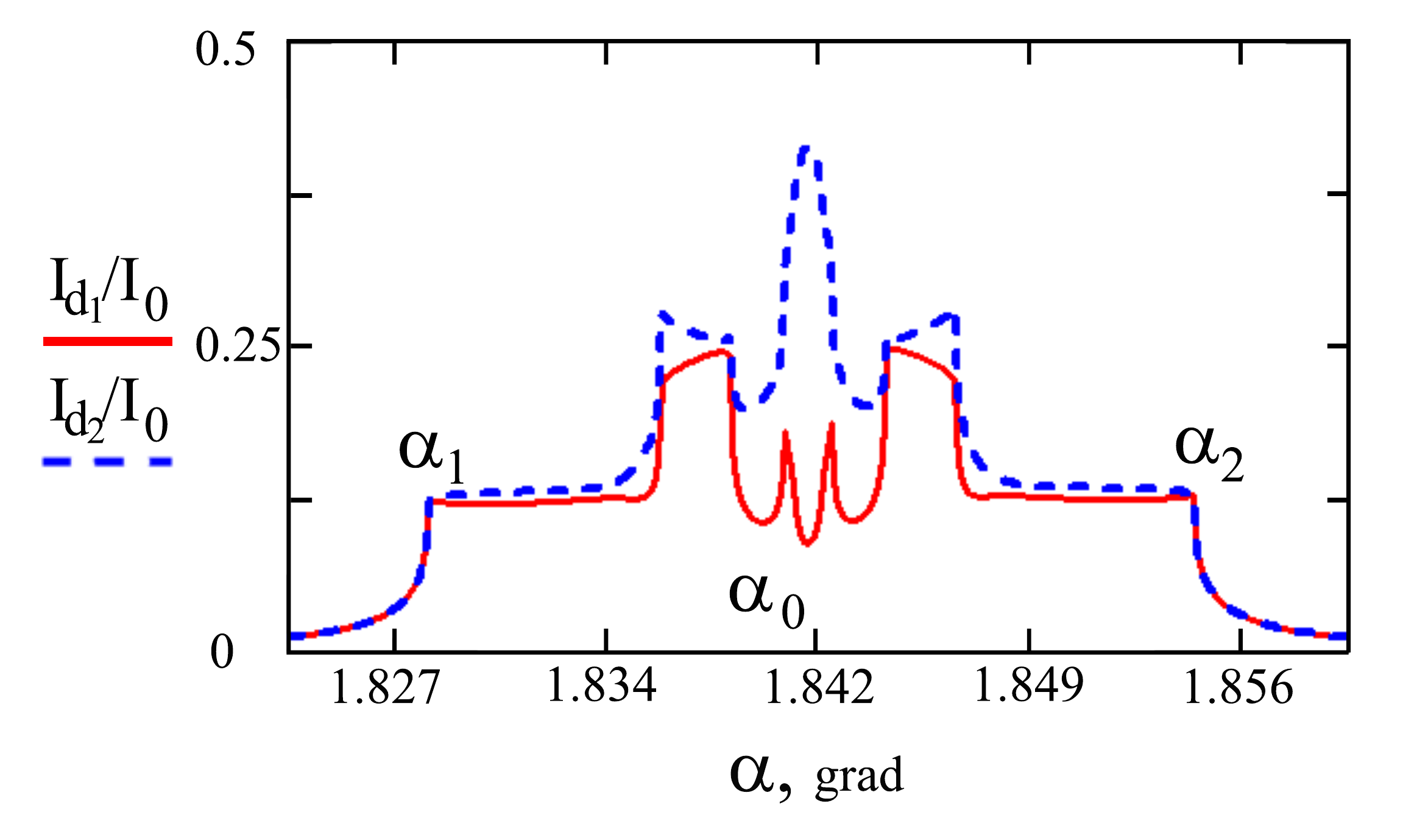}%
\caption{\label{Fig2} (Color online) Intensity of the waves diffracted in directions $\vec{k}_{d_{1,2}}$ (see Fig.~\ref{Fig1}) as a function of the glancing angle $\alpha$. The solid line corresponds to neutrons traveling to the right from the incident plane, and the dashed line corresponds to neutrons traveling to the left. $I_0$ is the intensity of an incident beam.}
\end{figure}

Figure~\ref{Fig2} shows the dependence of the diffracted beams intensities $I_{d_{1,2}}$ on the glancing angle $\alpha$ at the most symmetric case $\beta=0$.  The region between the angles $\alpha_1$ and $\alpha_2$ corresponds to the Darwin plateau. The main feature here is the asymmetry of diffraction, $I_{d_1}\ne I_{d_2}$. Changing the ratio of magnetic potentials leads to reshaping of fine structure of the plateau with the main feature staying the same. The asymmetry can be attributed to the "Hall effect" for neutrons. The sign of the asymmetry is changing by switching of the magnetic parameters $B_0$ and $B_z$. This switching can be achieved by the reversal of an external magnetic field from $\vec{B}$ to $-\vec{B}$. The asymmetry does not depend on the sign of $B_{\perp}$.

Recently it was demonstrated that the "Hall effect" for neutrons appears due to the time reversal symmetry breaking for spin 1/2 particles moving in a non-coplanar magnetic structure~\cite{Ud2013}. In dynamical theory of neutron diffraction the time reversal symmetry breaking occurs when at least two reciprocal vectors satisfy the Bragg condition. This condition follows from the following considerations. When the Bragg condition is valid for one reciprocal vector only, neutrons inside a crystal are described by Eq.~\ref{Eq_TwoWaveSyst} We assume that the vectors $\vec{B}_0$ and $\vec{B}_g$ are located in an (x,y)-plane. Applying time reversal operator to the system Eqs.~(\ref{Eq_TwoWaveSyst}) the vectors $\vec{B}_0$ and $\vec{B}_g$ are replaced by $-\vec{B}_0$ and $-\vec{B}_g$. Since both the vectors belong to the (x,y)-plane this switching can be compensated by rotation of spin coordinates by the angle $\pi$ around the z axis, $\hat{R}_z^{\pi}=i\sigma_z$. The spin rotation does not affect the unpolarized incident beam. Therefore, intensities of all scattered waves remains unchanged after the time reversal, and Eqs.~(\ref{Eq_TwoWaveSyst}) can be considered as obeying the time reversal symmetry.

When two reciprocal vectors satisfy the Bragg condition, neutrons are described by the equations~(\ref{Eq_GenSys}) containing four vectorial magnetic Fourier harmonics. These vectors do not belong to the same plane. We assume that three of these vectors form a left-hand system. The time reversal operation turns the left-hand system into the right one. There is no operator, which can compensate time reversal since a handedness may not be switched by the rotation of coordinate system. Therefore, the time reversal symmetry is broken leading to the Hall effect and asymmetrical diffraction of neutrons in the MnSi crystal.

Figure~\ref{Fig2} shows that there is no scattering asymmetry beyond the Darwin plateau ($\alpha <\alpha_1$ and $\alpha >\alpha_2$). In these regions the Bragg condition is not valid for two reciprocal vectors and the "Hall" effect is zero.

Magnitude of the "Hall effect" depends on the incident angle $\alpha$. Averaging over incident angles reduces the left-right asymmetry. The asymmetry $2(I_{d_1}-I_{d_2})/(I_{d_1}+I_{d_2})$ is about 80\% in the region close to angle $\alpha_0$. After averaging over the angles in the region between $\alpha_1$ and $\alpha_2$ the asymmetry decreases by 20\%. Thus, an incident beam divergence in the (x,z) plane should be comparable with the Darwin plateau angular width $\alpha_2-\alpha_1=0.03^o$ to observe the strong neutron "Hall effect" in experiment. We mention that the same requirement is imposed for the beam divergence in the (y,z) plane.

\begin{figure}[h]
\includegraphics[width=1\columnwidth, keepaspectratio=true]{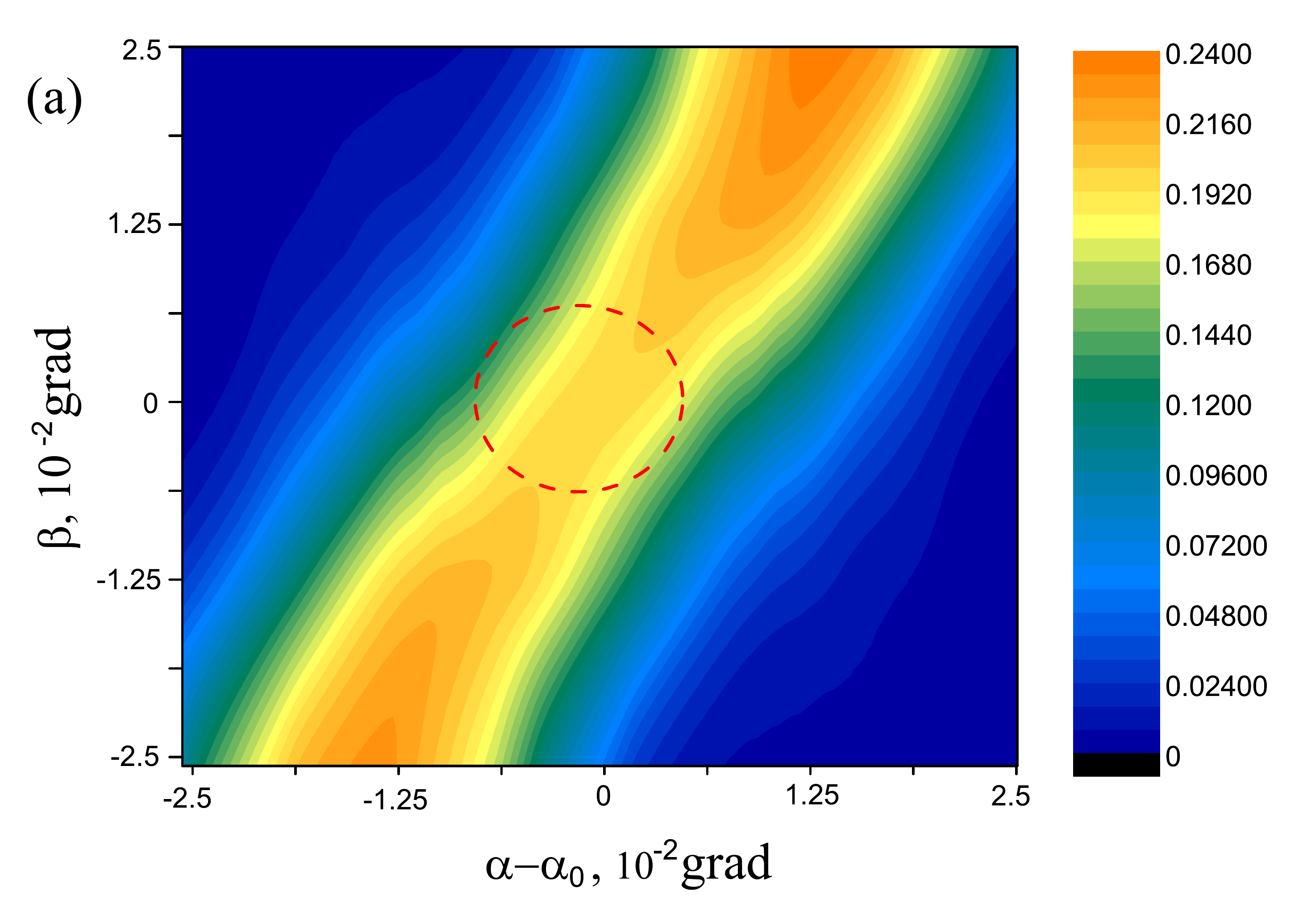} \\ \includegraphics[width=1\columnwidth, keepaspectratio=true]{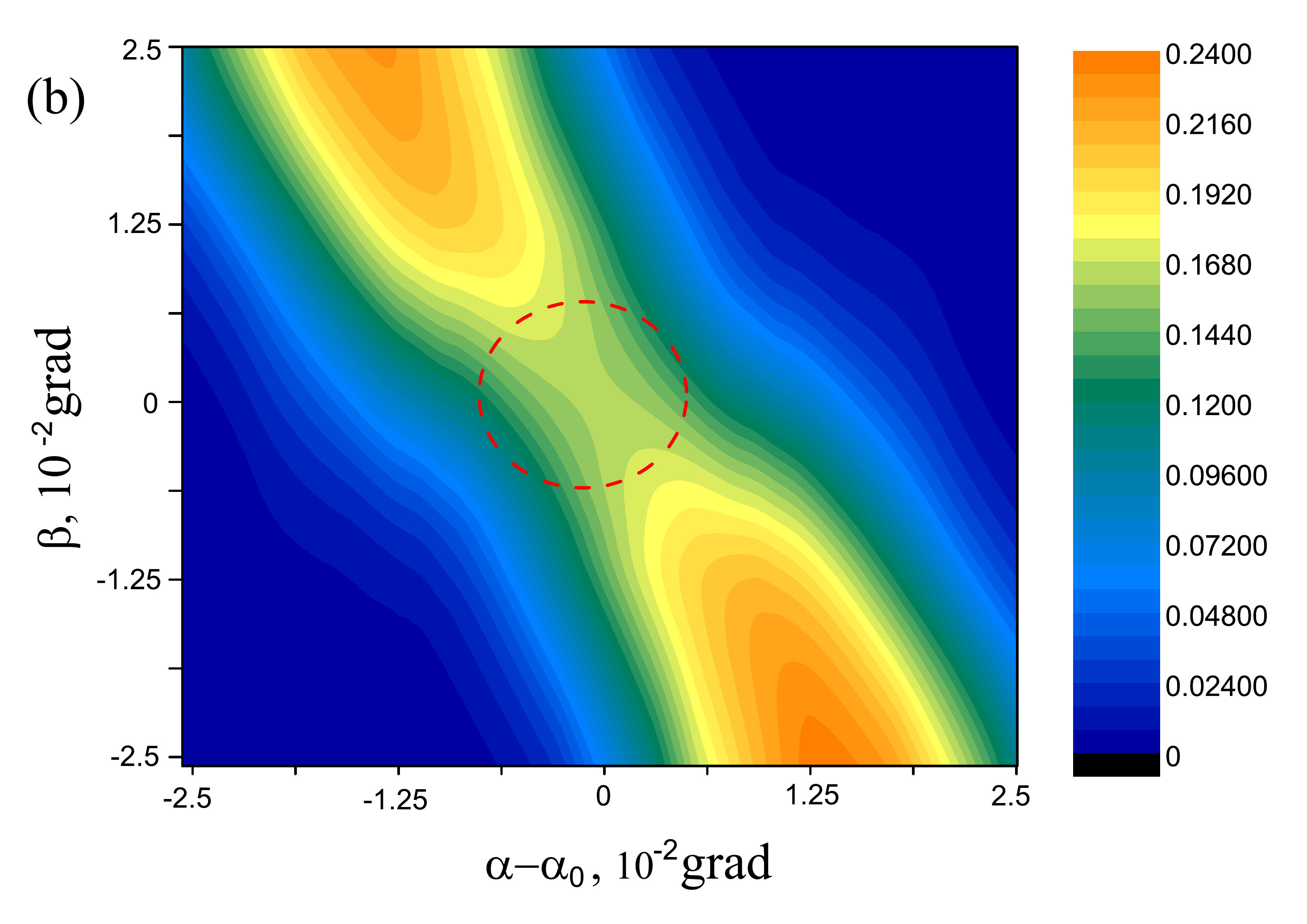}
\caption{\label{Fig_DifAv} The dependence of the averaged diffracted intensity on angles $\alpha$ and $\beta$. $\alpha_0=1.842$ grad. (a) and (b) correspond to diffraction directions $\vec{k}_{d_1}$ and  $\vec{k}_{d_2}$. Averaging was done over both angles $\alpha$ and $\beta$. Averaging area is 0.02$^o$. Circles show the regions in which two reciprocal vectors satisfy the Bragg condition and three wave approximation has to be used. In these regions skew scattering appear leading to difference between diffraction peaks $\vec{k}_{d_1}$ and  $\vec{k}_{d_2}$ and violation of the relation $I_{d_1}(\alpha,\beta)=I_{d_2}(\alpha,-\beta)$.}
\end{figure}

Fig.~\ref{Fig_DifAv} demonstrates the dependencies of the intensity of diffracted peaks averaged over angles $\alpha$ and $\beta$. Averaging window is the same for the both angles $\Delta\alpha=\Delta\beta=$0.02$^o$. Such an angular resolution is achievable in experiment. One can easily see the difference of diffracted intensities in the central regions of Fig.~\ref{Fig_DifAv}(a) and Fig.~\ref{Fig_DifAv}(b). The difference $\Delta I(\alpha, \beta)$ averaged over angles $\alpha$ and $\beta$ is presented in Fig.~\ref{Fig_DeltaDif}(b). After averaging the difference magnitude decreases to 20\% (compare with Fig.~\ref{Fig_DeltaDif}(a)). The effect is rather high and may be observed experimentally.

We now discuss the experimental data of neutron diffraction in the A phase of MnSi~\cite{Pfleiderer2009Sc}. Experimental geometry in this work is almost the same as discussed here (compare figure 1(B) of Ref.~\cite{Pfleiderer2009Sc} and figure S2 of a supporting online material of Ref.~\cite{Pfleiderer2009Sc} with Fig.~\ref{Fig1}). The only difference is that in the experiment a neutron beam falls on the back surface of MnSi crystal ((x,y) plane) instead of the (x,z) plane. However, it is known that a neutron reflection by the MnSi surface is rather weak and can be neglected. Let us consider the data on the second sample studied in Ref.~\cite{Pfleiderer2009Sc}. The sample Skyrmion lattice is oriented symmetrically with respect to the laboratory coordinate system similar to the present paper. Detailed experimental data on this sample is presented in the supporting online material of Ref.~\cite{Pfleiderer2009Sc}. According to Ref.~\cite{Pfleiderer2009Sc} the diffraction peaks 6 and 7 in the figure 2(E) correspond to the diffracted beams $\vec{k}_{d_1}$ and $\vec{k}_{d_2}$ of the current study. Figure S3(B) of the supporting online material shows that an angular position of the peaks is in a good agreement with our estimates and is approximately 2 degrees. However, there is almost no difference between the peaks 6 and 7, in contrast to our prediction. This is a consequence of rather high beam divergence which can be estimated as $\delta\alpha=1^o$. The divergence $\delta\alpha$ is 50 times larger than the angular region where the Hall effect is significant.

To conclude, we studied a neutron diffraction by the A phase of MnSi using dynamical theory of diffraction beyond the two wave approximation when two reciprocal lattice vectors satisfy the Bragg condition. New effect appears beyond the two wave approximation, namely the skew scattering of neutrons. In the symmetric case (when the incident neutron wave vector lies in the same plane as the Skyrmion axis) diffraction pattern is asymmetric with respect to the incident plane. The asymmetry sign can be changed by the reversal of an external magnetic field. The effect can be considered as the Hall effect for neutrons. The magnitude of the effect reaches 80\% in the small region of solid angles corresponding to the Darwin plateau. In the case of asymmetric orientation of the incident beam the skew scattering also leads to violating of neutron diffraction pattern symmetry.

The financial support by the "Dynasty" foundation is
gratefully acknowledged. AF acknowledges the support of The Ministry of education and science of the Russian Federation and RFBR. Authors would like to thank I.S. Beloborodov and Yu.N. Khaidukov for useful discussions.

\appendix

\section{Calculation procedure\label{Sec:App}}

The calculation procedure of the intensity of diffracted beams is the same as was used in Ref.~\onlinecite{Ud2013}. Solution of Eq.~\ref{Eq_GenSys} gives 12 eigenwaves $\Psi_{\vec{k}}$ inside the MnSi crystal. Each wave can be presented in the following way
\begin{equation}\label{WF}
\Psi_{\vec{k}}=\Psi_0 e^{i(\vec{k}\cdot\vec{r})}+\Psi_r e^{i((\vec{k}+\vec{g}_1)\cdot\vec{r})}+\Psi_l e^{i((\vec{k}+\vec{g}_2)\cdot\vec{r})},
\end{equation}
with $\Psi_{0,l,r}$ being dependent on $\vec{k}$, obviously. Six of them go outward the MnSi surface and the other six go toward the surface. We take into account only the outgoing waves. Full wave function of a neutron inside the crystal can be written as
\begin{equation} \label{TWF1}
\Psi^{\pm}_{cr}=\sum_{j=1}^{6} C^{\pm}_j\Psi_{\vec{k}_j}.
\end{equation}
The vectors $\vec{k}_j$ are found from Eq.~\ref{Eq_GenSys} taking into account the energy conservation law $\epsilon(\vec{k}_j)=|\vec{k}_{inc}|^2\hbar^2/(2m_n)$ (where $\epsilon(\vec{k}_j)$ is the neutron energy in the crystal, $\vec{k}_{inc}$ is the incident neutron wavevector) and the momentum conservation law, which takes the following form in the considered system $\vec{k}_j=\vec{k}_{inc}+\Delta_j\vec{n}$ ($\vec{n}$ is the MnSi surface inward normal). \cite{Stassis1974} Values $\Delta_j$ are the solutions of the dispersion equation appearing from the system (\ref{Eq_GenSys}). Superscript $\pm$ corresponds the spin state of the incident neutron beam.

Outside the crystal the neutron wave function is the combination of incident, reflected and two diffracted waves.

\begin{equation} \label{WFout}
\begin{split}
\Psi^{\pm}_{out}&=\hat{S}^{\pm}e^{i((\vec{k}^{||}_{inc}+\vec{k}^{\perp}_{inc})\cdot\vec{r})}+\hat{R}^{\pm}e^{i((\vec{k}^{||}_{inc}-\vec{k}^{\perp}_{inc})\cdot\vec{r})}+
\\&+\hat{D}^{\pm}_1e^{i((\vec{k}^{||}_{inc}+\vec{g}^{||}_1-\Gamma_1\vec{n})\cdot\vec{r})}+\hat{D}^{\pm}_2e^{i((\vec{k}^{||}_{inc}+\vec{g}^{||}_2-\Gamma_2\vec{n})\cdot\vec{r})}.
\end{split}
\end{equation}

Here superscripts "$||$" and "$\perp$" indicate parallel and perpendicular (with respect to the interface) components of the wave vectors. In contrast to the coefficients $C_i$ the quantities $\hat{R}$, $\hat{D_1}$, and $\hat{D_2}$ are the spinors. $\hat{S}^+=(1 0)^T$, $\hat{S}^-=(0 1)^T$. The coefficient $\Gamma_1$ is determined by the energy conservation law, which can be written as $|\vec{k}^{||}_{inc}+\vec{g}^{||}_1-\Gamma_1\vec{n}|=|\vec{k}_{inc}|$. The similar restriction is applied for calculation of $\Gamma_2$, namely, $|\vec{k}^{||}_{inc}+\vec{g}^{||}_2-\Gamma_2\vec{n}|=|\vec{k}_{inc}|$.

Wave functions Eq.~\ref{WF} and Eq.~\ref{WFout} obey the boundary conditions, \cite{Stassis1974}

\begin{equation}\label{RefEQ_3}
\left\{\begin{array}{l}
{\hat{S}^{\pm}+\hat{R}^{\pm}=\sum_{j=1}^{6}C^{\pm}_j\Psi^{k_j}_0}\\
{|\vec{k}^{\perp}_{inc}|(\hat{S}^{\pm}-\hat{R}^{\pm})=\sum_{j=1}^{6}C^{\pm}_j\Psi^{k_j}_0|\vec{k}^{\perp}_j|}\\
{\hat{D}^{\pm}_1=\sum_{j=1}^{6}C^{\pm}_j\Psi^{k_j}_r}\\
{-\Gamma_1\hat{D}^{\pm}_1=\sum_{j=1}^{6}C^{\pm}_j\Psi^{k_j}_r|\vec{k}^{\perp}_j|}\\
{\hat{D}^{\pm}_2=\sum_{j=1}^{6}C^{\pm}_j\Psi^{k_j}_l}\\
{-\Gamma_2\hat{D}^{\pm}_2=\sum_{j=1}^{6}C^{\pm}_j\Psi^{k_j}_l|\vec{k}^{\perp}_j|}
\end{array}\right.
\end{equation}

~
~

These equations give the amplitudes of reflected and diffracted waves. Reflected waves have intensity several orders of magnitude less than the diffracted waves in the region of Darwin plateau. Therefore we can neglect them. To calculate the intensity of the beams diffracted in the $\vec{k}_{d_1}$ and $\vec{k}_{d_2}$ directions for the unpolarized incident neutrons we introduce the following intensities $I_{diff}^{++}$, $I_{diff}^{+-}$, $I_{diff}^{-+}$, $I_{diff}^{--}$. Here superscripts denote the spin states of incident and diffracted neutrons. The intensities of the diffracted waves were summed over the spin states and also normalized to the sum of the intensities of incident beams with different spin orientations:

\begin{equation} \label{GrindEQ__28_}
I_{diff}=(I_{diff}^{++}+I_{diff}^{-+}+I_{diff}^{+-}+I_{diff}^{--})/2I_{0}.
\end{equation}

~
~

\bibliography{NeutronsRef}

\end{document}